\documentclass[12pt]{article}

\usepackage{graphics}
\begin{document}
\begin{titlepage}
\centerline{\large \bf Photoabsorption spectrum of the Xe@C$_{60}$ endohedral fullerene}  
\vglue .1truein
\centerline{Zhifan Chen and Alfred Z. Msezane}
\centerline{\it Department of Physics and CTSPS, Clark Atlanta University, Atlanta, GA 30314 USA} 
\centerline{}
\centerline{}
\centerline{\bf ABSTRACT}
    Photoabsorption spectrum of the Xe@C$_{60}$ endohedral fullerene has been studied 
using the time-dependent-density-functional-theory (TDDFT), which represents the
dynamical polarizability of an interacting electron system by an off-diagonal 
matrix element of the resolvent of the Liouvilliam superoperator and solves the 
problem with the Lanczos algorithm. The method has been tested with the   
photoabsorption cross sections of
the free Xe atom and C$_{60}$ fullerene.  
The result for the Xe@C$_{60}$ 
confirms the three main peaks observed in the recent measurement in the energy region of the 
Xe 4$d$ giant resonance and
demonstrates the underestimation of the photoionization cross section by the experiment.  
It is suggested to use the current theoretical result around 22 eV to
calibrate the experimental data.  

\centerline{} 

\end{titlepage}

\section{Introduction}
    Photoabsorption and photoionization cross sections of the Xe@C$_{60}$ molecule 
have been 
studied extensively by many groups both theoretically \cite{r1,r2,r3,r4,r5} and 
experimentally \cite{r6}.
Puska and Nieminen \cite{r1} modeled for the first time the C$_{60}$ molecule using a 
spherical shell with an attractive potential and used a jellium shell to calculate the
4d photoabsorption cross section of a Xe atom inside the C$_{60}$ molecule. 
Their result demonstrated some oscillations in the cross section curve. Amusia 
and
co-workers \cite{r2,r3,r4} used a $\delta$-like potential for the C$_{60}$
shell and the random-phase approximation with exchange (RPAE) method to describe  
the photoionization process. Their results indicate that the giant resonance is
distorted substantially, exhibiting four prominent peaks [3]. 
Dolmatov and Manson \cite{r5} showed that 
two of those 
peaks disappear if the finite thickness of the C$_{60}$ cage is utilized. 
However, their result is not supported by the recent 
experiment \cite{r6}.
The experimental data demonstrate strong enhancement of the photoionization cross
sction as well as an oscillatory structure of the amplitude. 
The complicated experimental resonance structure is also quite different from 
the calculation of the time-dependent density-functional theory (TDDFT) \cite{r7}, 
which considers all the valence electrons to form a delocalized charged cloud and 
treats the residual ion-core as a classical jellium shell. 
Recently, Chen and Msezane performed a calculation on
the confinement resonance of the Xe@C$_{60}$. They  
solved the Schr\"odinger equation in three separate regions, $r<r_i$, $r_i<r<r_o$ and
$r>r_o$ using both regular and irregular solutions, where $r_i$ and $r_o$ are, 
respectively the inner and outer radii of the C$_{60}$ spherical shell. 
The RPAE method was utilized to
describe the photoionization process. 
This method has been successfully used in the study of the photoionization of the 
encapsulated  
Ce$^{3+}$ \cite{r8,r9} and Xe$^+$ \cite{r10} ions in C$_{60}$. For the photoionization of the
Xe@C$_{60}$ molecule they obtained an 
improved agreement   
between theory and experiment \cite{r11}. 
Although the photoionization of the Xe@C$_{60}$ molecule has been studied using various  
theoretical models,
however, a first principle calculation is still needed.  

   In this paper we have performed a 
time-dependent density-functional-theory (TDDFT) calculation for the photoabsorption
spectrum of the Xe@C$_{60}$. The calculations of the optical
spectra are carried out in the linear response formalism. 
The most widespread calculation of the electronic excitation in TDDFT 
is the Casida's approach \cite{t1}, which 
reformulated the calculation of the response function into a generalized Hermitian 
eigenvalue problem.
However, this method is not suitable for calculating the spectra in a
broad energy range. 
Recently a method has been developed, which determines the dipole susceptibility 
based on the application of the 
Lanczos algorithms to the first order expansion of the time-dependent Kohn-Sham equation
in the linear response regime \cite{t2,t3,t4}. Here the 
absorption spectrum is calculated from the imaginary part of the dipole
susceptibility. The method avoids the 
explicit calculation of the unoccupied states of the ground state and allows the 
entire spectrum of a molecule to be computed.  
This new implementation of the TDDFT has been used in the present Xe@C$_{60}$ calculation.

\section{Method and results}
    We used the DMol$_3$ software of density-functional theory to determine the 
structures of the C$_{60}$ and Xe@C$_{60}$ and the TDDFT to evaluate the photoabsorption 
spectra. 

    The TDDFT method was used first to calculate the photoabsorption spectra 
of the C$_{60}$ fullerene. 
The C$_{60}$ geometry optimization was performed using the generalized gradient 
approximation (GGA) to the density-functional theory, with PBE 
exchange-correlation functional \cite{r15} 
along with all electron double numerical plus polarization (DNP) basis sets as 
implemented in the DMol$_3$ package \cite{r16}. The optimization of atomic 
positions proceeded until the change in energy was less than 5$\times$10$^{-4}$
eV and the forces were less than 0.01 eV/\AA. The resultant structure has 
the length of the shorter bond (hexagon-hexagon) 1.392 {\AA} and the longer 
bond (hexagon-pentagon) 1.448 {\AA}. These value are in good agreement
with the x-ray data value of 1.391 {\AA} for the shorter bond and 1.455
{\AA} for the longer bond \cite{r17}. The structure was  
validated using a plane-wave approach as implemented in the QUANTUM-ESPRESSO 
\cite{r18}.
A supercell of 18 {\AA} was constructed to eliminate the interactions 
among cells. An ultrasoft
pseudopotential of the Rappe-Rabe-Kaxiras-Joannopoulos (RRKJ) \cite{r19} 
type and a kinetic energy 
cutoff of 408 eV for the wave function and 2448 eV for the densities and potentials 
were employed in a standard ground 
state DFT calculation,  
yielding the Kohn-Sham eigenvalues and eigenvectors for all the occupied states.
The dynamical polarizability of an interacting electron system is represented 
as an appropriate off-diagonal matrix element of the resolvent of 
a Liouvillian superoperator. The resolvent of the Liouvillian is evaluated
through an algorithm based on the Lanczos method \cite{t2,t3,t4}. 

    Fig 1 presents the photoabsorption spectrum for the C$_{60}$ after 3000 Lanczos 
iterations. Solid curve is from our TDDFT calculation and the dashed line reprsents
the measurements \cite{r20,r21,r22}. The first three peaks at 3.48, 4.36, 
and 5.36 eV agree very well with
the other TDDFT and local orbitals calculation \cite{r23}. Their results show 
the peaks at  
3.5, 4.4, and 5.4 eV. The C$_{60}$ fullerene has a giant resonance,  
which corresponds to a collective oscillation of
delocalized electrons relative to the ionic cage. The peak around 5-7 eV 
is the $\pi$ plasmon resonances. The largest   
peak around 22 eV is 
related to the $\sigma$ surface plasmon resonance, which was first predicted by
the linear response theory \cite{r24}. The second collective resonance near 30-40 eV 
is associated with a dipole-excited volume plasmon \cite{r25}.
Our solid curve shows the second volume plasmon resonance predicted by the
experiment \cite{r25}. 
The over all excellent agreement has been obtained between our TDDFT result and the 
measurements.
\begin{figure}
\resizebox{0.85\columnwidth}{!}{%
  \includegraphics{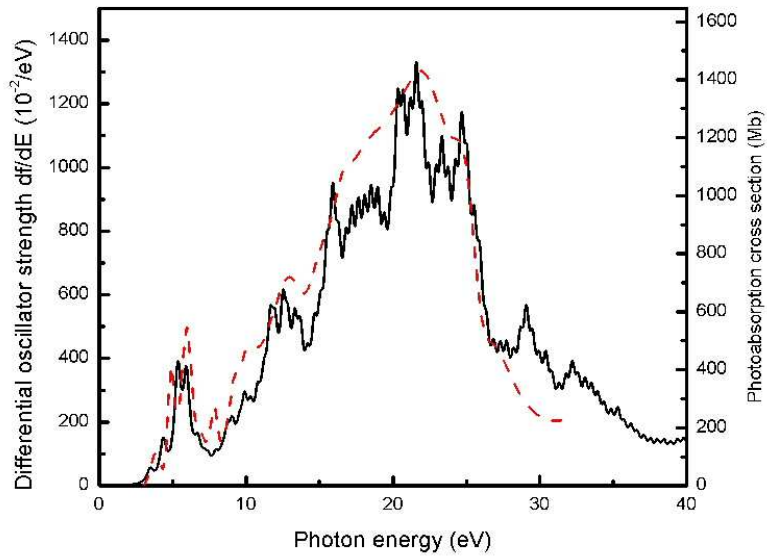}
}
\caption{
(Color online) Photoabsorption spectrum of C$_{60}$. Solid curve represents the 
result of the TDDFPT
calculation. Dashed line are the experimental data \cite{r20,r21,r22} 
}
\label{fig:1}
\end{figure}
   
    The TDDFT method using the Liouville-Lanczos approach has also been used 
to evaluate the photoabsorption spectra for a free Xe atom. In order to study the 4d 
giant resonance of a Xe atom, an ultrasoft pseudopotential,
which includes 4d electrons as valence electrons has been created using the ld1.x 
program of the QUANTRUM-ESPRESSO with the Xe ground state as the reference 
electron configuration.
The pseudopotential has multiple-projector with PBE of GGA
exchange-correlation functional and includes nonlinear core corrections.
Fig. 2 shows the calculated photoabsorption spectra. Solid line is the 
calculated absolute differential oscillator strength of 
the photoabsorption of the Xenon atom in the energy region of 8.43 eV ($5p$ threshold) 
-140 eV. Triangles are the experimental data obtained by Chan {\it et al} \cite{r26} 
using a low-resolution
spectrometer. 
Solid squares and solid circles are respectively, the measured photoionization 
cross sections 
of the $4d$ subshell of Xenon atom  
by Holland {\it et al} \cite{r27} and 
K\"{a}mmerling {\it et al} \cite{r28}.
Note that the unit of photoionization cross section on the right vertical coordinate
is the Mb, which is related to the differential oscillator strength through, 
1.0975 Mb=1$\times$10$^{-2}$ differential oscillator strength unit (1/eV)
\cite{r29}. Our calculation 
describes those measured data very well.
\begin{figure}
\resizebox{0.85\columnwidth}{!}{%
  \includegraphics{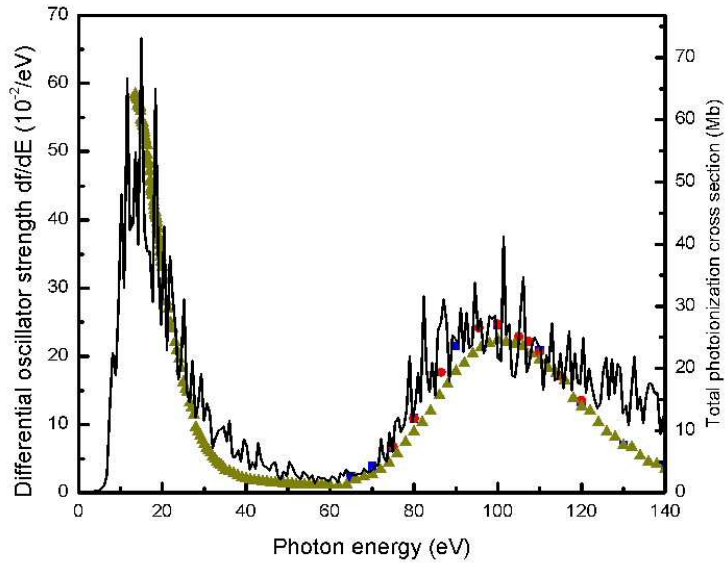}
}
\caption{
(Color online) Photoabsorption spectrum of free Xe atom. Solid curve is 
our TDDFT results. 
Triangles, solid squares and solid circles are respectively, from
the measurements of \cite{r26}, \cite{r27} and \cite{r28}.
}
\label{fig:2}
\end{figure}
   
   The structure of the Xe@C$_{60}$ was obtained by adding a Xe atom into the center of
the C$_{60}$ fullerene and then performing the same geometry optimization as that of 
the C$_{60}$. The kinetic energy cutoff of 1020 eV for the wave function and 
4080 eV for the density
were used in the ground state calculation.
Fig. 3 shows the photoabsorption spectrum of the Xe@C$_{60}$. In the energy range
of the Xe 4$d$ giant resonance, we can find several small wiggles which do not exist in
the C$_{60}$ case. 
\begin{figure}
\resizebox{0.85\columnwidth}{!}{%
  \includegraphics{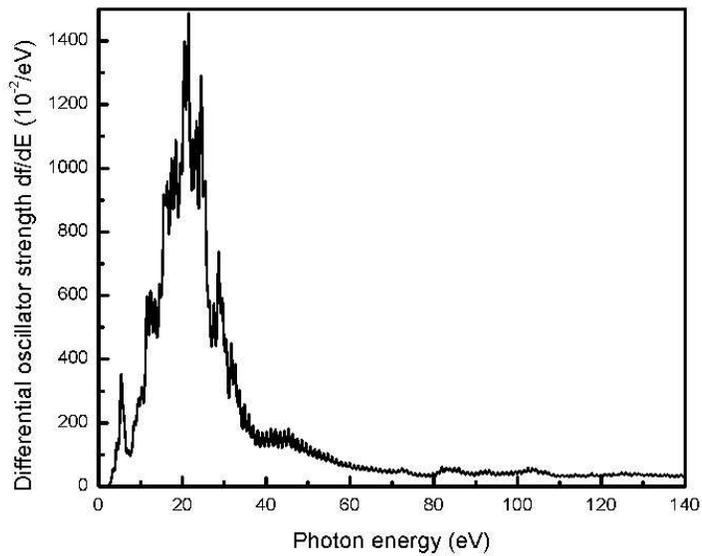}
}
\caption{
Photoabsorption spectrum of Xe@C$_{60}$
}
\label{fig:3}
\end{figure}
   The enlarged details of these wiggles are presented in Fig. 4. Fig. 4 also 
shows the convergence situations of the 4d resonance. Dotted line, 
dashed line and solid line represent,
respectively the 1500, 3500 and 4500 Lanczos iterations.  
\begin{figure}
\resizebox{0.85\columnwidth}{!}{%
  \includegraphics{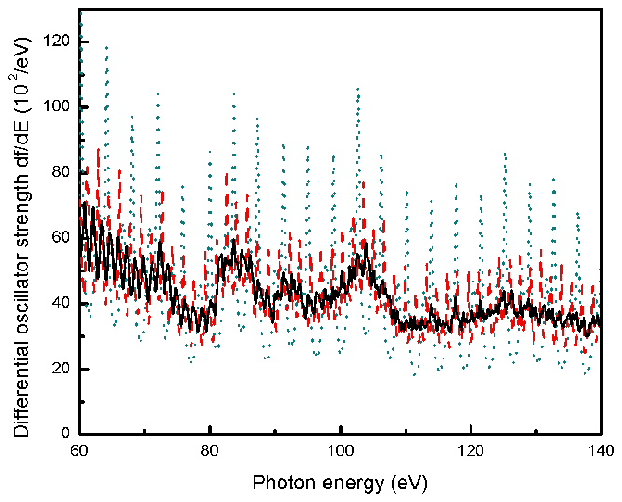}
}
\caption{
(Color online) Comparison between the spectra at different iterations. Dotted, 
dashed and solid
curves represent, respectively 1500, 3500 and 4500 iterations.
}
\label{fig:4}
\end{figure}
    
   Fig. 5 is the comparison of our TDDFT calculation with the experiment
in the energy range of the Xe $4d$ giant resonance. Solid circles with the error bars
are from experiment \cite{r6}. Solid curve represents our calculation. 
Since the experiment underestimates
the photoionization cross sections we have enlarged the measured cross sections
by a factor of seven and shifted them upwards by 0.25/eV.  
The calculated Xe@C$_{60}$ curve is  
shifted to the right by
7 eV to match the three main peaks of the experiment. Our calculation confirms the
three main peaks in the measurement and demonstrates that the experiment has underestimated
the photoionization cross section of Xe@C$_{60}$.  
We recommend the use of the TDDFT result at the low energy region to calibrate the 
measurement.For example, at 21.47 eV the photoionization cross section from 
TDDFT is 1633 Mb. 
\begin{figure}
\resizebox{0.85\columnwidth}{!}{%
  \includegraphics{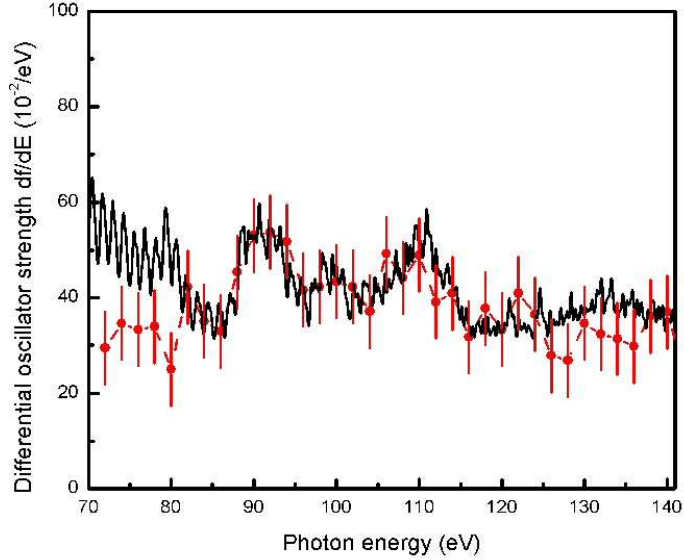}
}
\caption{
(Color online) Comparison between the TDDFT calculation and the experimental data.
Solid circles with error bars are taken from Ref. [6]. Solid curve is our TDDFT
calculation. Experimental data have been enlarged seven times and shifted upwards by
0.25/eV. The calculated curve has been shifted to the right by 7 eV.  
}
\label{fig:5}
\end{figure}

\section{Conclusion}
   In conclusion, photoabsorption spectra for the Xe@C$_{60}$ molecule has been 
studied using the TDDFT method. The result confirms the three main peaks observed
in the experiment \cite{r6}. The calculation also reveals the underestimation of 
the photoionization cross section by the measurement. It is suggested to
use the TDDFT result around 22 eV to calibrate the measurement. 

\bigskip
\centerline{\bf ACKNOWLEDGMENTS}
\bigskip

   This work was supported by the U.S. DOE, Division of Chemical Sciences,
Geosciences and Biosciences, Office of Basic Energy Sciences, Office
of Energy Research, AFOSR and Army Research Office (Grant W911NF-11-1-0194).
Calculations used Kraken system of the National Institute for Computational
Science, The University of Tennessee. 

\end{document}